# A Novel Design of Capacitive Plasmonic Near Field Transducer

Tianxiang Du [1, 2], David E Laughlin[1, 2, 3], and Jian-Gang (Jimmy) Zhu[1, 2, 3]

[1]Data Storage Systems Center, Carnegie Mellon University, Pittsburgh, PA 15213 USA

[2]Department of Materials Science and Engineering, Carnegie Mellon University, Pittsburgh, PA 15213 USA

[3]Department of Electrical and Computer Engineering, Carnegie Mellon University, Pittsburgh, PA 15213 USA

Corresponding author: Tianxiang Du

Email: tianxiangdu@gmail.com

## Abstract

A near field transducer (NFT) is a key photonics component in heat assisted magnetic recording (HAMR) for the localized heating of the magnetic medium. In this work, we present a novel NFT design through capacitive coupling. In our design, tapered metal bars separated by thin dielectric materials with gap distance G are used to create the plasmonic resonance and focus the electromagnetic field. The design is motivated by the intention to improve thermal stability, which can be achieved through segmentation using thermally stable dielectric material between the plasmonic metal bars. Using COMSOL Multiphysics software, the performance of this capacitive-coupled NFT is systematically modeled. It is shown that the electromagnetic field could gradually be focused through the tapering towards the air bearing surface (ABS). In addition, the focusing effect could be enhanced with a smaller NFT peg size at the resonant wavelength. The material selection for the NFT tip material will be discussed to further address the thermal stability of the device. In conclusion, this capacitive-coupled NFT with dielectric separation gaps and tapering yields an enhanced |E| field intensity at the tip with the potential for an enhanced material thermal stability. Such a design can also exhibit applications in other energy delivery systems as well as plasmonic waveguides and sensors.

# I. Introduction

In heat assisted magnetic recording (HAMR), a near field transducer (NFT) is used to generate a very localized intensive electric field, heating the metallic magnetic grains above the Curie Temperature through the Joule heating so that a write transducer is able to apply the magnetic field to the magnetic grains for the HAMR writing process. [1][2][3] Current industry NFT designs use a solid piece of plasmonic metal with certain geometric shape for resonance mode matching. For a common industry design, a NFT peg with tens of nanometer dimension is used to exert the electric field to the media metallic grains at optical frequencies. [4] However, materials with excellent plasmonic properties such as gold, are subject to deformation under an elevated temperature due to the highly mobile grain boundaries, resulting in the reliability challenges that the HAMR technology has been facing ever since its inception. [3][5][6]

Recently, several research groups reported that dielectric materials such as $Al_2O_3$ or $TiO_2$ coatings could significantly increase the mechanical and thermal stability of the embedded materials. For example, Zazpe *et al* reported that the $TiO_2$ nanotube with a 10 nm thick $Al_2O_3$ coating could maintain its microstructure stability after annealing at 870 °C for 1 hour. [7][8] In our previous work, we also proposed a capacitive nanocomposite structure with the Au metal bars separated by the thin dielectric materials. [9] The idea of the capacitive design is the utilization of thermally stable dielectric materials to physically constrain the deformation of Au metal bars. In the sections to follow, we will further optimize our capacitive nanocomposite design through tapering towards the air bearing surface (ABS).

# II. The NFT Design

Figure 1 shows our 3D tapering design of the NFT where the yellow part denotes the gold, and the dark blue part denotes the dielectric material. The light can be coupled through a dielectric waveguide that leads an incident electric field in the plane towards the NFT to excite the structure. The NFT is formed by an array of gold bars with a fixed dimension, $L_{bar}$, along the direction of plasmonic wave propagation. Each bar is separated by a thin dielectric gap with a gap distance G (G is around 10 nm), resulting in the capacitive coupling between adjacent bars. The Au metal bars are gradually tapered along the transverse direction with a tapering angle α, resulting in a final NFT tip size, $W_{tip}$. By matching the wavelength of the incident light, the intrinsic material plasmonic characteristics, and the geometric parameters of the NFT, a resonance coupling between the incident light and the plasmonic oscillation inside the NFT can be achieved. An enhanced electromagnetic field can then be generated on the surface of NFT tip, forming a hot spot across the air bearing surface (ABS).

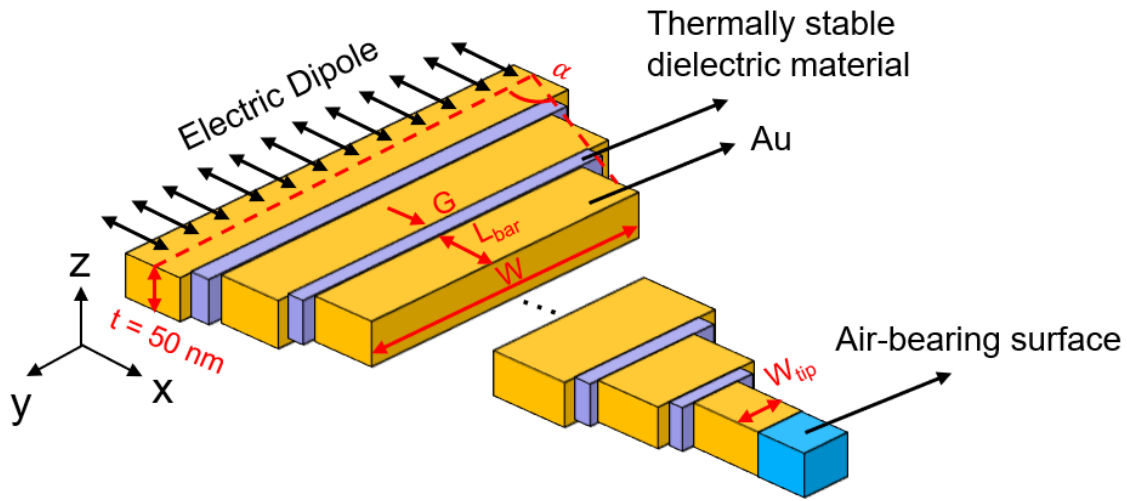

Figure 1: Schematic drawing of the tapering NFT structure with capacitive coupling.

The electromagnetic simulation for the 3D capacitive nanostructure utilized the Electromagnetic Waves, Frequency Domain interface (emw) in the COMSOL Multiphysics radio frequency (RF) module. The capacitive nanostructure was composed of Au metal bars embedded in the dielectric medium. A series of electric dipole sources with inter-spacing as 25 nm were placed at the left end of the nanostructure with x component as -0.001 A•m to end-excite the nanostructure. The distance between the dipole sources and the left end of the 1st bar was kept constant at 50 nm. The bar thickness t was kept constant as 50 nm along the z direction and the refractive index of the dielectric medium was set as $n_{diel} = 1.5$. For the capacitive structure, the gap distance G was kept as 10 nm and the bar length $L_{bar}$ was fixed at 100 nm . The tapering angle α could be varied to change the final tip width $W_{tip}$ while the total length of the structure was fixed with the exact length value specified in each following section. The whole system was surrounded by a perfectly matched layer (PML) with a thickness of 300 nm to absorb the electromagnetic wave reaching the boundary of the dielectric medium. A second order scattering boundary condition was implemented at the outer boundary of PML to further prevent any wave reflection back into the system. The Au dielectric constants as a function of the sweeping wavelength was implemented based on P. B. Johnson and R. W. Christy's work. [10] For all the calculation below, the electric field amplitude was taken in the center of the dielectric gap along the propagation direction.

## III. Results and Discussions

For the following calculation, a systematic modeling analysis will first be presented with a focus on the wave propagation modes in the 3D untapered structure. The focusing effect of the electromagnetic wave will then be illustrated with a gradual tapered width W to demonstrate the significance of the tapering design. The influence factors of the focusing effect will be analyzed, followed by the discussion of NFT tip material selection.

### A. Electric Field Propagation for 3D Untapered NFT

In this part, the E field propagation as a function of the bar width W was calculated at a fixed free wavelength of $\lambda_{free}$ = 950 nm for the untapered capacitive nanostructure. The objective of the calculation is to understand the effect of the bar width W regarding the wave propagation since the width direction is another degree of freedom in determining the properties of the plasmonic wave. Here we set the dielectric gap G = 10 nm, and the total number of Au gratings was as 21. All other parameters remained unchanged as mentioned above.

Figure 2 (a) shows the general structure of the untapered capacitive nanostructure. Figure 2 (b) shows the average $|E|^2$ along the last gap by integrating $|E|^2$ along the gap center and divided by the bar width W. Figure 2 (c)-(f) shows the $|E|^2$ propagation for different width W, corresponding to the position 1-4 in Figure 2 (b), representing either constructive interference (Figure 2 (c), (e), (f)) or destructive interference (Figure 2 (d)). It can be observed that different wave propagation modes could appear due to the boundary confinement of different bar width W. Modes with multiple peaks can form when the bar width W is larger than 200 nm while only a single peak mode can be observed if the bar width W is smaller than 200 nm. The observation is significant because a final tapered NFT structure needs to be developed with bar width W getting tapered

towards the air bearing surface. Therefore, it is necessary to understand the influence of bar width W on the propagation of the plasmonic wave.

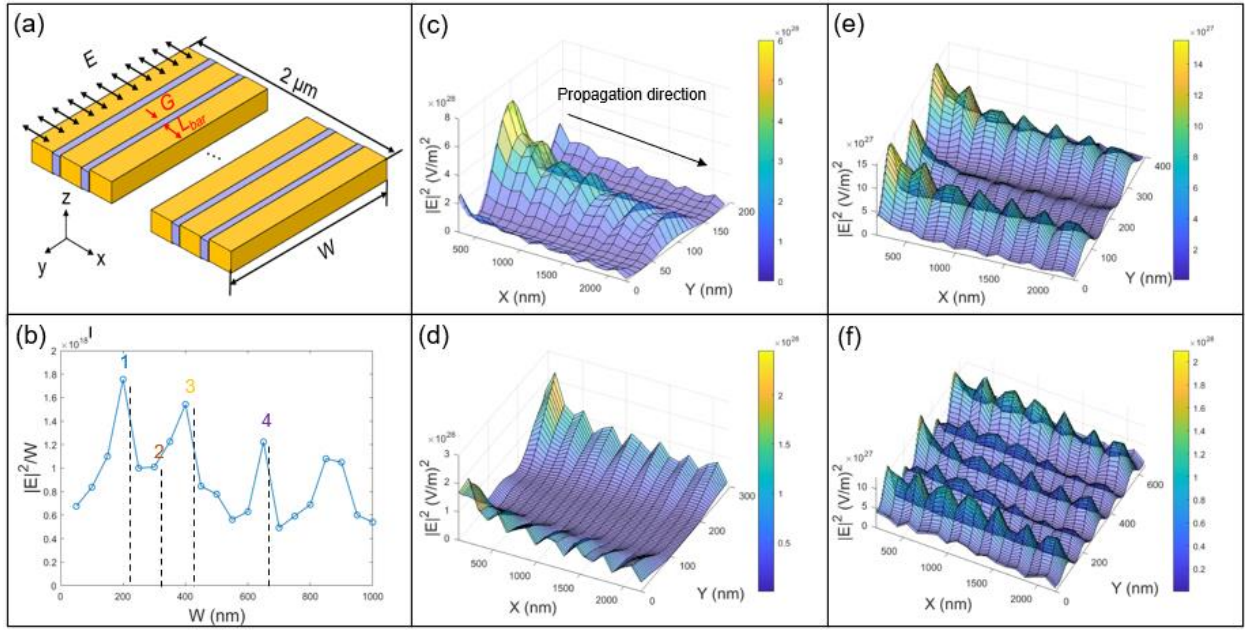

Figure 2: Plasmonic wave propagation for 3D untapered structure. (a) illustrates the general untapered geometry, (b) shows the average $|E|^2$ vs bar with W at $\lambda_{free} = 950$ nm, (c)-(f) shows the $|E|^2$ propagation for different bar width W, the $|E|^2$ is extracted only in the center of the dielectric gap.

To further analyze the electromagnetic wave mode with respect to the bar width W, we plotted the $|E|^2$ along the y direction for the last gap center (Figure 3). It can be seen that as the wave propagation changes from the multiple peak mode to the single peak mode, the amplitude of the peak also increases. Therefore, a single peak mode is preferred to deliver the energy. In addition, Figure 3 shows the potential feasibility of tapering the Au segments towards the air bearing surface, where the plasmonic wave can evolve from a multiple peak mode to a single peak, thereby focusing and enhancing the amplitude of the electromagnetic field. A detailed analysis of the tapered NFT design will be presented in the next section.

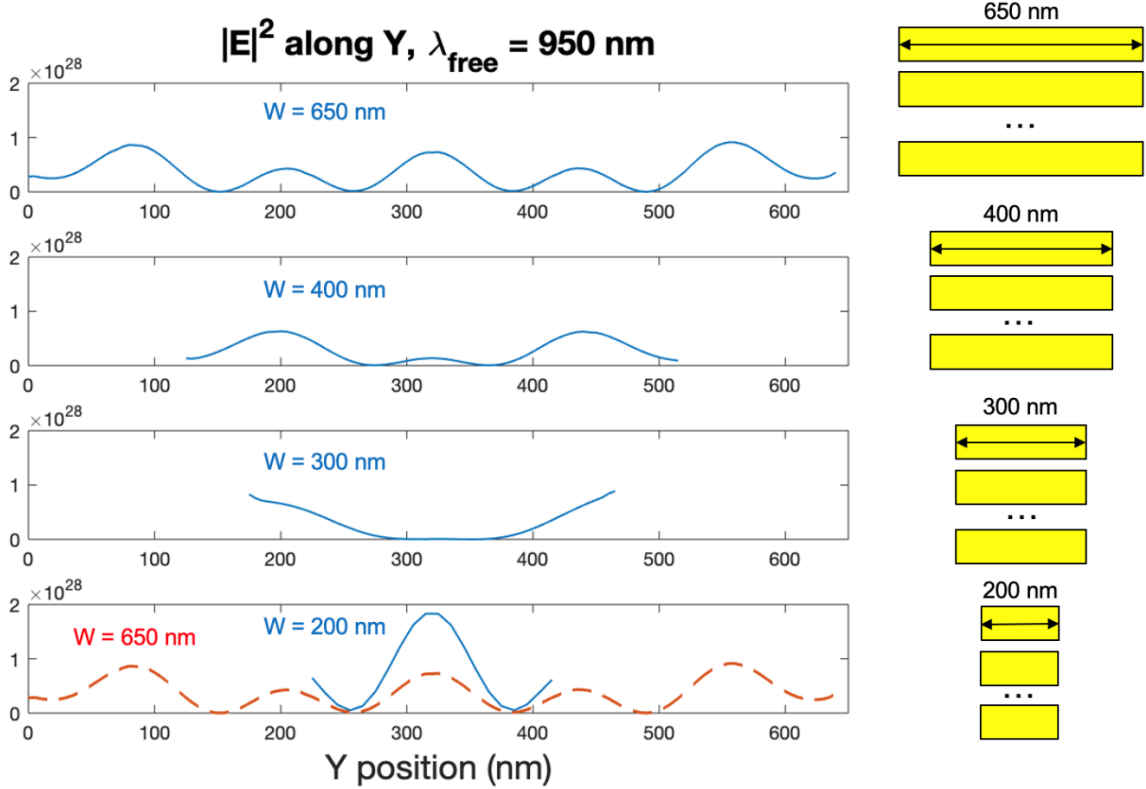

Figure 3: $|E|^2$ along the y direction for the last dielectric gap, the last subplot shows the comparison between the $|E|^2$ amplitude for W = 650 nm and W = 200 nm.

## B. Electric Field Propagation for 3D Tapered NFT

In this section we compared the E field propagation for the 3D untapered structure and the tapered structure. Figure 4 shows the geometry and the $|E|^2$ distribution for the two geometries at a free wavelength of $\lambda_{free}$ = 1180 nm, which is a resonant wavelength for the tapered structure in Figure 4 (a). Both the configurations contained N = 14 Au bars before the last piece with a dielectric gap distance G = 10 nm and the first bar width W = 900 nm. For the tapered structure, the tapering angle α was set as 73.5°, resulting in a final tip width $W_{tip}$ = 25 nm.

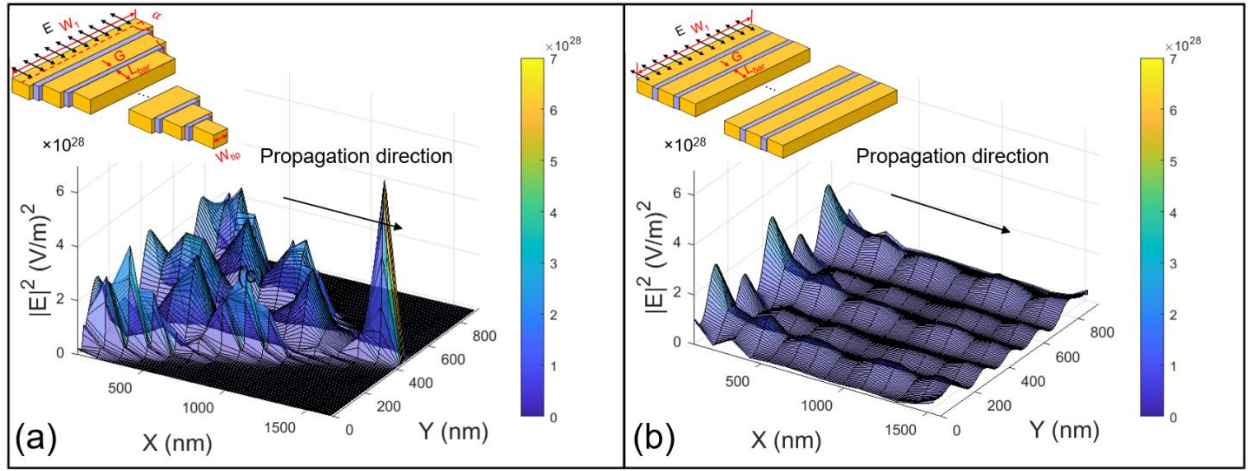

Figure 4: $|E|^2$ propagation comparison between the tapered (a) and the untapered structure (b) at $\lambda_{free} = 1180$ nm. The inset in (a) and (b) shows the geometry of the tapered and the untapered structure, respectively.

It can be observed that the |E| field propagation for both the configurations starts with the similar multiple peak mode. For the tapered structure, the plasmonic wave propagates and gradually gets focused, resulting in a peak intensity much higher than the beginning (Figure 4 (a)). For the untapered structure, the plasmonic wave propagates and gradually decays while forming a standing wave pattern along the propagation (Figure 4 (b)), which is similar to the standing wave pattern in the 2D capacitive structure. [9] Therefore, the tapered design of the NFT exhibits a focusing effect and increases the $|E|^2$ intensity delivered to the magnetic media.

## C. Factors Influencing the Focusing Effect

### 1. Influence of the NFT Tip Width $W_{tip}$

To explore how the NFT tip width affects the focusing effect, we kept the same total length and varied the tapering angle α so that the NFT tip width $W_{tip}$ could be changed. Figure 5 (a) shows the case of the $|E|^2$ along the last gap when we varied the tapering angle α from 73.5° to 75°, resulting in the increase of the tip width $W_{tip}$ from 25 nm to 106 nm. It is found that the areas under the three curves are similar, indicating a similar total energy output. However, with a smaller tapering angle α (thus smaller tip width $W_{tip}$), we could achieve a larger $|E|^2$ peak intensity and therefore, enhancing the focusing effect. Figure 5 (a) also shows that the Full-Width-Half-Maximum (FWHM) of the energy profile curve becomes smaller with a small tip size $W_{tip}$, demonstrating that an enhanced and concentrated electromagnetic field can be applied in the HAMR heating process. Figure 5 (b) shows the peak intensity $I_p$ as a function of the tip width $W_{tip}$ = 25 nm, 53 nm, 80 nm, 106 nm, 160 nm and 214 nm, corresponding to the tapering angle α as 73.5°, 74°, 74.5°, 75°, 76° and 77°, respectively. It can be observed that the intensity increases monotonically with a smaller NFT tip width $W_{tip}$, which further validates our conclusion.

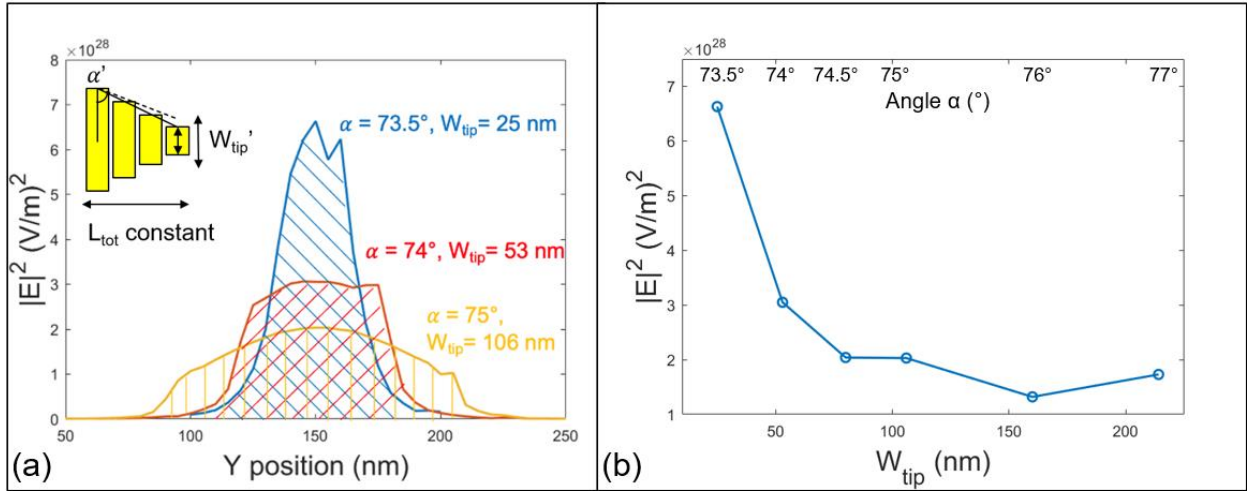

Figure 5: Focusing effect vs NFT tip width $W_{tip}$ at $\lambda_{free}$ = 1180 nm. (a) shows the $|E|^2$ profile along the last gap for structures with different tapering angle α and tip width $W_{tip}$, (b) shows the $|E|^2$ peak intensity as a function of the NFT tip width $W_{tip}$.

## 2. Influence of the Excitation Wavelength $\lambda_{free}$

Here we investigate the focusing effect regarding the excitation free wavelength $\lambda_{free}$. Figure 6 (a) shows the average $|E|^2$ in the last gap vs the scanning free wavelength $\lambda_{free}$, indicating 1180 nm as the resonant wavelength and 1040 nm as the off-resonant wavelength. Figure 6 (b) and (c) show the resonant and the off resonant $|E|^2$ profile along the last gap for $W_{tip}$ = 25 nm, 53 nm, and 106 nm, respectively, demonstrating that the focusing effect is wavelength dependent. For a specific geometry, it is critical to excite the structure by applying the resonant wavelength so that the $|E|^2$ field amplitude could be maximized. It is worth noting that the excitation of the capacitive nanocomposite structure relies on the proper design of a dielectric waveguide, which is not included in this work.

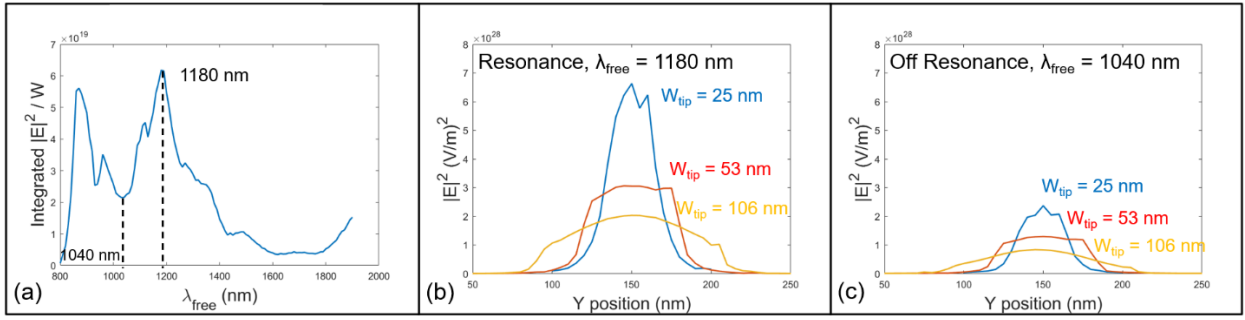

Figure 6: Focusing effect vs free wavelength $\lambda_{free}$, (a) shows the average $|E|^2$ vs $\lambda_{free}$ for the last gap with a tip width of $W_{tip}$ = 25 nm. (b) and (c) shows the $|E|^2$ along the last gap at $\lambda_{free}$ = 1180 nm and 1040 nm, respectively.

## D. Effects of the Tip Material Selection

As mentioned earlier, the NFT tip experiences the most intensive electromagnetic field during its operation, primarily attributable to the focusing effect. Therefore, to prevent the deformation of the NFT tip caused by the highly mobile gold grain boundaries [3], various alternative material systems could be considered for enhancing the thermal stability. Figure 7 shows the NFT efficiency versus the dielectric constants of the tip material with the configurations of 2-segment NFT and 6-segment NFT as insets, where the yellow, blue, orange and grey color denotes the Au, thermally stable dielectric material, tip material and HAMR media layer, respectively. The length of the 1st and 2nd bar (tip) in the 2-segment NFT is 750 nm and 150 nm, where the tip material can be replaced with another material M. For the 6-segment NFT, each segment has a length of $L_{bar}$ = 150 nm. Both two configurations have the same initial width as W = 900 nm and the same dielectric gap distance G = 5 nm. The tapering angle α was set as 63.4° and all other parameters were kept the same as in section II. The HAMR media layer has a stack FePt (magnetic medium, 10 nm)/MgO (10 nm)/Cr (50 nm) with the spacing between the NFT and the media as 5 nm. The dielectric constants for FePt, MgO and Cr were fixed at -12.46-29.16*i, 2.96 and -1.26-23.96*i, respectively. [11][12]

The ratio of the power dissipated in the FePt media over the NFT was applied to demonstrate the NFT heating efficiency, as shown in the equation below. To characterize the optimum operating wavelength, we selected the maximum NFT efficiency out of the scanning wavelength from 850 nm to 1900 nm for each of the configurations.

$$\text{NFT Efficiency} = \frac{P_{\text{Media}}}{P_{\text{NFT}}}$$

It can be observed in Figure 7 (a) and (b) that a small imaginary part $\varepsilon_{\text{img (tip)}}$ and a large real part $\varepsilon_{\text{real (tip)}}$ of the tip dielectric constants can contribute to an increase in the NFT efficiency, since a lower optical loss and a larger E field enhancement can be achieved. Such a tip material could be fabricated by doping Au with a small amount of hardening material such as Ir. [13] In addition, the 6-segment NFT configuration shows a decreasing NFT efficiency compared to the 2-segment NFT due to the optical loss introduced by the dielectric gap. [9] However, the increasing amount of the dielectric material added in the 6-segment NFT system could potentially enhance the thermal stability of the device compared to the 2-segment NFT design.

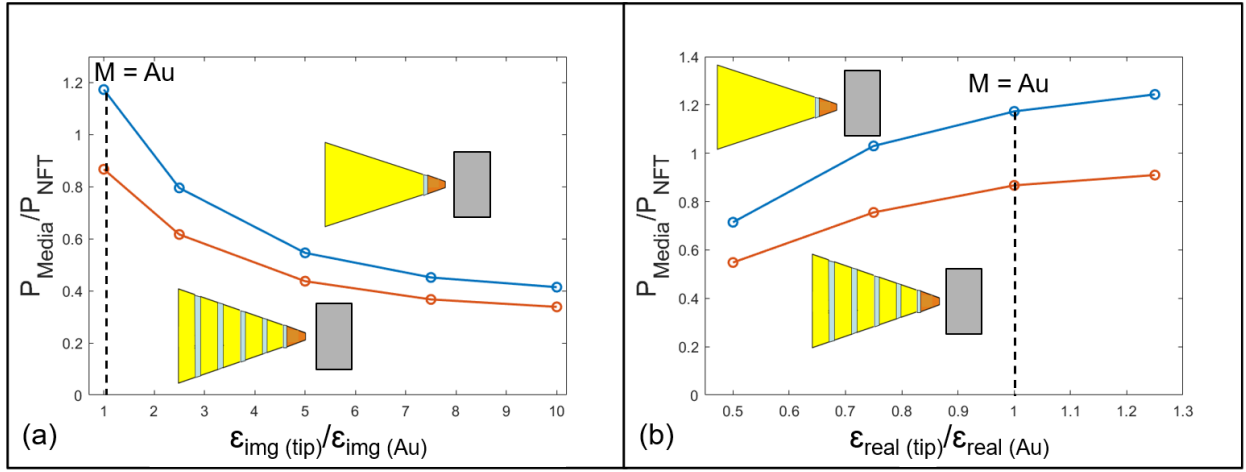

Figure 7: NFT efficiency vs tip material properties. (a) shows the efficiency vs the imaginary part of the tip material and (b) shows the efficiency vs the real part of the tip material. The insets show the corresponding configuration related to the curve.

## IV. The Optimal NFT Geometry

Based on all the analysis before, we now propose our optimal NFT geometry design, as shown in Figure 8. The NFT consists of 14 Au segments with the length of each segment $L_{\text{bar}} = 100$ nm. The Au segment starts with an initial width $W_1 = 900$ nm with a tapering angle $\alpha = 73.5°$, resulting in a final NFT tip size $W_{\text{tip}} = 25$ nm. The dielectric material between has a refractive index $n_{\text{gap}} = 1.5$ (corresponding to the refractive index of $SiO_2$) with a dielectric gap distance G =

5-10 nm. The whole structure has a thickness t = 50 nm and needs to be excited by an electric field E along the x direction. Under a resonant free wavelength $\lambda_{free}$ = 1180 nm, the structure can output a maximum electric field amplitude $|E|^2$ of $6.6 \times 10^{28}$ (V/m)$^2$.

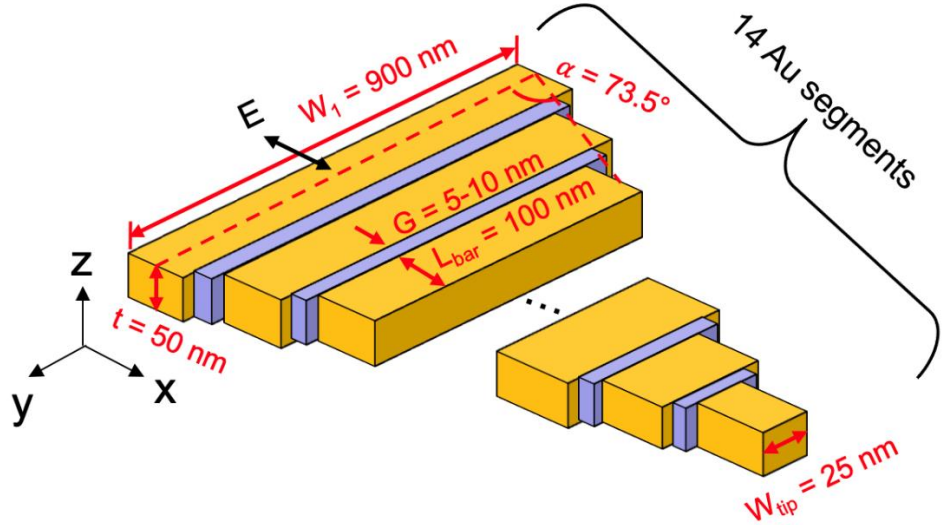

Figure 8: Illustration of the optimal NFT geometry design.

It should be noted that the NFT tip material can be replaced to address the thermal stability of the device. According to our discussion in section III D, doping Au with a few percent of Ir for the tip material is a promising solution, in which case we can achieve a large real part dielectric constant $\varepsilon_{real}$ and a relatively small imaginary part dielectric constant $\varepsilon_{img}$, thereby maintaining an acceptable NFT heating efficiency.

## V. Conclusions

In this paper, we proposed a tapered capacitive NFT design with an array of gold bars separated by the dielectric gaps, forming a grating structure with the tapering towards the air bearing surface (ABS). Simulation shows that different bar width W can form different wave propagation modes while a single mode is preferred to deliver the energy. Through a tapering design of the capacitive NFT, the electromagnetic field can be concentrated at the NFT tip to efficiently heat the magnetic media through Joule heating. In addition, the focusing effect is determined by the excitation wavelength $\lambda_{free}$ and the NFT tip width $W_{tip}$. To further address the thermal stability of the device, different tip material system should be considered with an emphasis on a large negative $\varepsilon_{\text{real (tip)}}$ and a small $\varepsilon_{\text{img (tip)}}$. To conclude, the tapered capacitive NFT design addresses the energy delivery towards the magnetic media with a potential to enhance thermal stability of the device.